\documentclass[preprint,preprintnumbers,amsmath,amssymb]{revtex4}
\usepackage{amssymb,latexsym,amsmath}

\usepackage[activeacute,english]{babel}
\usepackage{fancyhdr}
\usepackage{textcomp}
\usepackage{amsfonts}
\usepackage{lscape}
\usepackage{graphicx}
\usepackage{subfigure}
\usepackage{graphics}
\usepackage{color}
\usepackage{array}

\usepackage[active]{srcltx}

\newcommand{\al}{\alpha}

\newcommand{\ep}{\epsilon}

\newcommand{\De}{\Delta}

\newcommand{\rar}{\rightarrow}

\begin{document}

\title{Two charges on plane in a magnetic field: III. $He^+$ ion}

\author{M.A.~Escobar-Ruiz}
\email{mauricio.escobar@nucleares.unam.mx}

\affiliation{Instituto de Ciencias Nucleares, Universidad Nacional
Aut\'onoma de M\'exico, Apartado Postal 70-543, 04510 M\'exico,
D.F., Mexico}

\date{\today}

\begin{abstract}
The $He^+$ ion on a plane subject to a constant magnetic field $B$ perpendicular to the plane is considered taking into account the finite nuclear mass. Factorization 
of eigenfunctions permits to reduce the four-dimensional problem to three-dimensional one. The ground state energy of the composite system is calculated in a wide range of
magnetic fields from $B=0.01$ up to $B=100$ a.u. and center-of-mass Pseudomomentum $K$ from $0$ to $1000$ a.u. using a variational approach.
The accuracy of calculations for $B = 0.1 $ a.u. is cross-checked in Lagrange-mesh method and not less than five significant figures are reproduced in energy. Similarly to the case of moving neutral system on the plane a phenomenon of a sharp change of
energy behavior as a function of $K$ for a certain critical $K_c$ but a fixed magnetic field occurs.

\end{abstract}

\pacs{31.15.Pf,31.10.+z,32.60.+i,97.10.Ld}

\maketitle

\begin{center}
\section*{Introduction}
\end{center}

\hskip 1cm

We study a two body Coulomb system on the plane subject to a constant magnetic field perpendicular to the plane. Main focus of this paper is on the charged system, in particular 
the $He^+$ ion.

\begin{figure}[htp]
\begin{center}
\includegraphics[width=4.0in,angle=0]{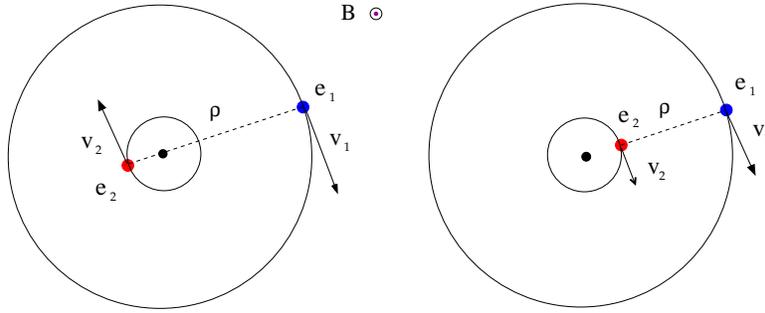}
\caption{For a $He^+$-like ion ($e_1=-2\,e_2$) a family of periodic circular trajectories occurs in which the
particles rotate with the same angular frequency on concentric circles with (left) relative phase $\pi$ or (right) zero relative phase. 
The relative distance between particles remains unchanged during the evolution.}
\label{T}
\end{center}
\end{figure}

\hskip 0.02cm
In classical mechanics a planar system of two Coulomb charges $(e_1,m_1)$ and $(e_2,m_2)$ in the presence of a constant magnetic field $B$ perpendicular to 
the plane is bounded for any value of magnetic field $B>0$ (see e.g. \cite{PM:2006}). In the case of particles with opposite charges ($e_1\,e_2<0$) for certain initial 
conditions special concentric (closed) trajectories occur (see \cite{ET:2013} and references therein), they are shown in Fig. \ref{T}. It manifests the appearance of 
extra conserved quantities specific for these trajectories (particular integrals of motion)

\vspace{0.7cm}

\hskip 1cm
In quantum mechanics, the appearance of particular integrals signals the existence of quasi-exactly-solvable solutions. In the cases 
of neutral atom ($e_1+e_2=0$) at rest and of quasi-equal charges $(\frac{e_1}{m_1}=\frac{e_2}{m_2})$ 
common analytical solutions of Hamiltonian and particular integrals emerge 
for certain discrete values of magnetic field strength \cite{ET-q:2013}-\cite{Turbiner:1994}. 
They also present the periodic circular trajectory Fig. \ref{T} (left). Therefore, the 
trajectories (Fig.\ref{T}) may indicate that the physically important case of the $He^+$ ion ($\alpha,e$) in magnetic field possesses analytical eigenfunctions. However, 
a single exact solution has not been found yet.

\hskip 1cm
It is worth describing the 3D case. In the three-dimensional case, quantum mechanical charged systems in magnetic field have been reviewed by Garstang \cite{Garstang} in the infinite nucleus mass
approximation. In the case of finite nuclear mass the CM motion cannot be separated from the relative motion. The investigation of the effects 
of CM motion on the properties of two-body systems in magnetic field started with a detailed mathematical study \cite{Avron}. 
To the best of our 
knowledge there was a single attempt \cite{Bezchastnov} to study the $He^+$ ion taking into account the finite mass effects. It was 
based on multiconfigurational
Hartree-Fock method and carried out
for the case of strong fields, $B\gtrsim 500$ a.u. 

In spite of the fact that there exists a number of properties which are common for two- and three-dimensional systems in a constant uniform magnetic field
a connection between two- and three-dimensional cases is unknown. The aim of the present work is not to study those common properties, but they will be mentioned below.

\hskip 1cm
For the planar quantum $He^+$ ion in a magnetic field perpendicular to the plane calculations of eigenfunctions are not available. Unlike the cases 
of neutral atom ($e_1+e_2=0$) at rest and of quasi-equal charges $(\frac{e_1}{m_1}=\frac{e_2}{m_2})$ the CM motion of the $He^+$ ion cannot be (pseudo)separated 
from the relative motion as well.
The situation gets complicated due to the absence of (particular) 
integrability \cite{Turbiner:PI}. Nevertheless, one component of
the conserved Pseudomomentum $\mathbf K$ found by Gor'kov \& Dzyaloshinskii \cite{GD:1967} for 3D neutral system remains integral 
for a planar charged systems. It allows us
to reduce this four-dimensional problem to a three-dimensional one.

\hskip 1cm
In a previous paper \cite{ET-gen:2013} an accurate variational solution, complementary to the exact solutions, for several low-lying states for both quasi-equal
charges $(\frac{e_1}{m_1}=\frac{e_2}{m_2})$ and neutral system at rest was given. The accuracy of obtained results was evaluated in a specially designed, convergent perturbation theory. 
In \cite{ET-genq:2013} the moving neutral system was considered.
By studying the ground state energy it was shown the stability of the system for all studied magnetic fields.

\hskip 1cm
The goal of the present paper, which is the natural continuation of \cite{ET-gen:2013}-\cite{ET-genq:2013}, is to perform a detailed study of the ground state
of the $He^+$ ion for different magnetic fields and Pseudomomentum,
checking its stability. It has to be emphasized that the variational functions are chosen to be also eigenfunctions of one component of Pseudomomentum.
It is worth mentioning that the $He^+$ ion in 3D is seen as an important system for astrophysics \cite{Neutron} for large values of the
magnetic field.

\hskip 1cm
We are going to employ a variational method with an optimization of the form of the vector potential (optimal gauge fixing)
constructing the trial function in such a way to combine a WKB expansion at large distances with perturbation theory expansion at small distances
near the minima of the potential into an interpolation \cite{Turbiner:1988-2010}.

\bigskip

\section{Generalities}

The Hamiltonian, which describes the planar $He^+$ ion, $(e_1=-e\,,m_1)$ and $(e_2=2e,\,m_2)$ , in a constant and uniform magnetic field ${\bf B}=B\,\hat {\bf z}$ perpendicular to the plane,
has the form
\begin{equation}
\begin{aligned}
{ {\hat H}} = &  \frac{{({\mathbf {\hat p}_1}+\frac{e}{c}\,{\mathbf A(\boldsymbol \rho_1)})}^2}{2\,m_1} + \frac{{({\mathbf {\hat p}_2}-\frac{2\,e}{c}\,{\mathbf A(\boldsymbol \rho_2)})}^2}{2\,m_2}
- \frac{2\,e^2}{\mid {\boldsymbol \rho}_1 - {\boldsymbol \rho}_2 \mid} + {\hat H}_{spin} \,,\qquad \boldsymbol{\rho}_{1,2} \in \Re^2\ ,
\label{Hcar}
\end{aligned}
\end{equation}
($e>0$) where $\hslash=\frac{1}{4\,\pi\, \ep_0}=1$, ${\mathbf {\hat p}}_{1,2}=-i\,\nabla_{1,2}$ is the
canonical momentum, ${\boldsymbol \rho}_{1,2}$ the position vector, $m_{1,2}$ the mass of the first (second) particle, respectively. 
$\mathbf A\in \Re^2$ is the vector potential which
corresponds to a constant magnetic field ${\bf B}$. The spin contribution ${\hat H}_{spin}= g\,(\mathbf s_1 + \mathbf s_2)\cdot \mathbf B$ is disregarded in the following
because its contribution is trivial.

It is easy to check that the total Pseudomomentum,
\begin{equation}
{\mathbf {\hat K}}\ \equiv \ \mathbf k_1 + \mathbf k_2 \ =\ \bigg[\mathbf {\hat p}_1  +  \frac{e}{c}\,\mathbf A(\boldsymbol \rho_1) - e\,{\mathbf B} \times {{\boldsymbol \rho}_{1}}\bigg] +\bigg[
(\mathbf {\hat p}_2  -  \frac{2\,e}{c}\,\mathbf A(\boldsymbol \rho_2) + 2\,e\,{\mathbf B} \times {{\boldsymbol \rho}_{2}}\bigg]   \ ,
\label{pseudo}
\end{equation}
is a gauge-independent integral of motion belonging to the plane, on where the dynamics is developed,
\[
      [{\mathbf {\hat K}} \ ,\ {\hat H}    ]\ =\ 0\ .
\]

For a single charged particle $q$ in a constant magnetic field the guiding center $\boldsymbol \rho_{{}_c}$, the center of the classical
trajectory, can be written in terms of the pseudomomentum $\boldsymbol \rho_{{}_c}=\frac{{\mathbf k_q}\times \mathbf B}{e\,B^2}$. For the two-body  neutral system, ${\mathbf {\hat K}}$ coincides (up to a unitary transformation) with the total canonical momentum and for $B=0$ becomes the total kinematic momentum \cite{ET:2013}. In general, the components of Pseudomomentum, closely connected to the phase
space symmetries of the underlying classical and quantum Hamiltonians, are the generators of the phase space translation group \cite{Avron}.

\bigskip

The following quantity is also an integral
\begin{equation}
\boldsymbol  {\hat \Omega} \ =\ \bigg[ {{\boldsymbol \rho}_{1}} \times ({\mathbf {\hat p}_1}+\frac{e}{c}\,{\mathbf A(\boldsymbol \rho_1)}) - \frac{e\,B}{2\,c}\rho_1^2 \, \hat {\bf z} \bigg]
  +  \bigg[  {{\boldsymbol \rho}_{2}} \times ({\mathbf {\hat p}_2}-\frac{2\,e}{c}\,{\mathbf A(\boldsymbol \rho_2)}) + \frac{e\,B}{c}\rho_2^2 \,\hat {\bf z}  \bigg]          \ ,
\label{Lz}
\end{equation}
$[\, \boldsymbol  {\hat \Omega}, \,  {\hat H} \,]=0$. The vector $\boldsymbol  {\hat \Omega}$ is perpendicular to the plane. In 
the symmetric gauge ($\xi=\frac{1}{2}$), $\boldsymbol  {\hat \Omega}$ becomes the total canonical angular momentum of the system.

It is easy to check that the operators $\mathbf {\hat K} = (\hat K_x,\,\hat K_y),\ \boldsymbol  {\hat \Omega}= \hat \Omega\,\hat {\bf z}$ obey the commutation relations
\begin{equation}
\begin{aligned}
&[ {\hat K}_x,\,{\hat K}_y ] = -e\,B\,,
\\ & [ {\hat \Omega} ,\,{\hat K}_x ] = {\hat K}_y\,,
\\ & [ {\hat \Omega},\,{\hat K}_y ] = -{\hat K}_x\ ,
\label{AlgebraInt}
\end{aligned}
\end{equation}
Hence, they span a noncommutative algebra. The problem is not completely integrable. The Casimir operator ${\cal {\hat C}}$ of this algebra is nothing but
\begin{equation}
{\cal {\hat C}}\ =\  {\hat K}_x^2+{\hat K}_y^2-\frac{2\,e\,B}{c}{\hat \Omega} \ .
\label{Casimir}
\end{equation}

It is clear that the integrals (\ref{AlgebraInt}) form a subset of those already present in the three-dimensional case \cite{Avron}.

Now, let us introduce on the plane cartesian coordinates and consider a certain one-parameter family of vector potentials corresponding to a constant magnetic field ${\bf B}=B\,\hat {\bf z}$ \cite{PRepts}

$$\mathbf A_{\mathbf r}  \equiv \mathbf A(\mathbf r)\,=\, B\,(( \xi -1 )\,y\,,\, \xi\, x,\,0)  \ , $$

where $0 \eqslantless \xi \eqslantless 1$ is a real parameter and $\mathbf r=(x, y)$. If $\xi=\frac{1}{2}$ we get the well-known
and widely used gauge which is called symmetric or circular. If $\xi=0 \ \text{or}\ 1$, we get the asymmetric or Landau gauge.

It is convenient to introduce center-of-mass (c.m.s.) coordinates
\begin{equation}
\begin{aligned}
&\mathbf R = \mu_1\, {\boldsymbol \rho}_1 + \mu_2\,{\boldsymbol \rho}_2 \ ,
\quad  {\boldsymbol \rho}= {\boldsymbol \rho}_1 - {\boldsymbol \rho}_2\ ,
\\ & \mathbf {\hat P} = {\mathbf {\hat p}}_1 + {\mathbf {\hat p}}_2 \ ,
\qquad \quad \, \, {\mathbf {\hat p}} = \mu_2\,{\mathbf {\hat p}}_1 -  \mu_1\,{\mathbf {\hat p}}_2\ ,
\end{aligned}
\label{CMvar}
\end{equation}
where $\mu_i=\frac{m_i}{M}$ is the ratio of the mass of the $i$th charge to the total mass of the system $M = m_1 + m_2$. In these coordinates
\begin{equation}
\mathbf {\hat K}  = \mathbf {\hat P}\ - \frac{e}{c}\,\mathbf A_{\bf R} + \frac{e}{c}\, \mathbf B \times \mathbf R
+ \frac{e\,\mu}{c}\,\mathbf A_{\boldsymbol \rho}  -  \frac{e\,\mu}{c}\,\mathbf B \times {\boldsymbol \rho}  \ ,
\label{pseudoR}
\end{equation}
\begin{equation}
\begin{aligned}
\boldsymbol {\hat \Omega}\  =\ & \bigg[ \mathbf R \times ({\mathbf {\hat P}}-\frac{e}{c}\,{\mathbf A_{\mathbf R}}) + \frac{e\,B}{2\,c}R^2 \, \hat {\bf z} \bigg]
  +  \bigg[ {\boldsymbol \rho}  \times ({\mathbf {\hat p}}-\frac{q_{\text w}}{c}\,{\mathbf A_{\boldsymbol \rho}}) + \frac{q_{\text w}\,B}{2\,c}\rho^2 \,\hat {\bf z}  \bigg]
  \\ & + \frac{2\,e\,\mu}{c}\,\mathbf R \times {\mathbf A_{\boldsymbol \rho}} -  \frac{e\,\mu\,B}{c} (\mathbf R \cdot {\boldsymbol \rho})\, \hat {\bf z} \ ,
\label{LzR}
\end{aligned}
\end{equation}

(cf. (\ref{pseudo}), (\ref{Lz})), where $\mu = \mu_2+2\,\mu_1$ and $q_{\rm{w}} \equiv  e\,(2\,\mu_1^2 - \mu_2^2)$  is an effective charge (weighted total charge).

Now, following \cite{ET-q:2013} we make a unitary transformation of the canonical momenta
\[
  U^{-1}\,{\mathbf {\hat P}}\, U \ = \ {\mathbf {\hat P}} +  \frac{e\,\mu}{c}\,\mathbf B \times {\boldsymbol \rho} - \frac{e\,\mu}{c}\,\mathbf A_{\boldsymbol \rho}    \quad , \quad
U^{-1}\,{\mathbf {\hat p}}\, U \ = \ {\mathbf {\hat p}}-  \frac{e\,\mu}{c}\,\mathbf B \times {\bf R} + \frac{e\,\mu}{c}\,\mathbf A_{\bf R}
\ ,
\]
with
\begin{equation}
\label{U}
 {U}\ =\ e^{i\,\frac{e\,\mu}{c}\,(  \mathbf B \times {\boldsymbol \rho}   -\mathbf A_{\boldsymbol \rho})\cdot \mathbf R} \ .
\end{equation}
Then, the unitary transformed Pseudomomentum reads
\begin{equation}
  {\mathbf K^{\prime}}\ =\ U^{-1}\,{\mathbf {\hat K}} \, U \ = \ \mathbf {\hat P}\ - \frac{e}{c}\,\mathbf A_{\bf R} + \frac{e}{c}\, \mathbf B \times \mathbf R \ ,
\label{KTrans}
\end{equation}
and coincides with the CM momentum of the whole, composite system, see (\ref{pseudo}).
The unitary transformed Hamiltonian (\ref{Hcar}) takes the form
\begin{equation}
\begin{aligned}
{\cal {\hat H}}^{\prime} & \ ={U}^{-1}\ {\cal {\hat H}}\ U  \ = \
{\cal {\hat H}}_{CM}(\mathbf {\hat P},\mathbf R, \boldsymbol \rho) +
{\cal {\hat H}}_{rel}(\mathbf { \hat p},\boldsymbol \rho)
\\& \ \equiv
 \  \bigg[  \frac{ {( \mathbf {\hat P}-e\,\mathbf A_{\mathbf R}+e\,\mu \,\mathbf B \times {\boldsymbol \rho} )}^2}{2\,M}  \bigg]
+ \bigg[      \frac{{({\mathbf {\hat p}}-q_\text{w}\,{\mathbf A_{\boldsymbol \rho}})}^2}{2\,m_{r}} -\frac{2\,e^2}{\rho}  \bigg]      \ ,
\label{H}
\end{aligned}
\end{equation}

It is evident, $[\, \mathbf {\hat K}^{\prime}, \, {\cal {\hat H}}^{\prime} \,]=0$. The eigenfunctions of  ${\cal {\hat H}}^{\prime}$ and ${\cal {\hat H}}$ are related
\begin{equation}
   \Psi^{\prime}\ =\ \Psi\ e^{ -i\,\frac{e\,\mu}{c}\,(  \mathbf B \times {\boldsymbol \rho}   -\mathbf A_{\boldsymbol \rho})\cdot \mathbf R}\ .
\label{psiprime}
\end{equation}

Unlike the neutral system, for a charged system the components of the Pseudomomentum ($\ref{KTrans}$) do not commute with each other, see (\ref{AlgebraInt}). Therefore,
the eigenfunctions of the corresponding Schr\"{o}dinger equation can not be chosen as simultaneous eigenfunctions of the Pseudomomentum, but of one of its components only.

\bigskip

Immediately, one can check that the eigenfunction of ${\hat K}_x^{\prime}$ has the form

\begin{equation}
 \Psi^{\prime}_{{}_{K}}(\mathbf R\,, \boldsymbol \rho)\ = \ \text{e}^{i\,(e\,B\,\xi\,Y + K)X}\,\psi_{{}_{K}}(\boldsymbol \rho,Y)    \ ,
\label{psik}
\end{equation}

where $\mathbf R = (X,Y)$, $\boldsymbol \rho=(x,y)$, $K$ is the eigenvalue and
$\psi_{{}_{K}}(\boldsymbol \rho,Y)$ depends on the relative coordinates $\boldsymbol \rho$ and $Y$.
The factor $\text{e}^{i\,(e\,B\,\xi\,Y + K)X}$ represents the only $X$-dependent part of the total wave function $\Psi^{\prime}_{{}_{K}}$.

\clearpage

Substituting $\Psi^{\prime}_{{}_{K}}$ into the Schr\"{o}dinger equation with ${\cal {\hat H}}^{\prime}$ we obtain the equation for $\psi_{{}_{K}}$

\begin{equation}
\begin{aligned}
&  \hat h\,\psi_{{}_{K}}\ =\ E\, \psi_{{}_{K}}\
 \\ & \hat  h \ \equiv
   \frac{ (- \partial^2_{{}_Y} - 2\,\imath\,B\,e\,\mu\,x\,\partial_{{}_Y} + e^2\,B^2\,Y^2 +2\,e\,B\,Y\,(K - B\,e\,\mu \,y))}{2\,M}
+      \frac{{({\mathbf {\hat p}}-q_\text{w}\,{\mathbf A_{\boldsymbol \rho}})}^2}{2\,m_{r}} +V_{eff}                             \ ,
\end{aligned}
\label{HKa}
\end{equation}

with an effective (gauge-invariant) potential-like term \cite{ET-genq:2013}

\begin{equation}
V_{eff}(x,y) \ =\   \frac{(B^2\,e^2\,\mu^2\,x^2 + {(K - B\,e\,\mu \,y)}^2)}{2\,M} - \frac{2\,e^2}{\rho}  \ .
\label{VeffTot}
\end{equation}

where $\partial_Y \equiv \frac{\partial}{\partial\,Y}$ and CM momentum $K$ plays a role of external parameter. The equation (\ref{HKa})
has some similarity with that of the 2D moving neutral system \cite{ET-genq:2013}. By making the substitutions $e\rightarrow e/\sqrt{2}$
and $\mu \rightarrow \sqrt{2}$ both equations coincide when the first term in r.h.s. of (\ref{HKa}) is absent. A similar gauge-invariant term has been encountered in 3D as well \cite{Schmelcher}.
The equation $\hat h\,\psi_{{}_{K}} = E\, \psi_{{}_{K}}$ (\ref{HKa}) is the basic equation we are going to study. An immediate observation is that
the CM and relative coordinates are not separated.
The problem is essentially three-dimensional and we arrive at the question how to solve it.
A simple idea that we are going to employ is to combine a WKB expansion at large
distances with perturbation theory near the minima of the potential (\ref{VeffTot}) into an interpolation.
The main practical goal of this paper is to construct such an approximation for the ground state
of the $H_e^+$ ion and then use it as variational trial function.

\section{The effective potential-like term and optimal gauge.}
\bigskip

\begin{figure}[htp]
\begin{center}
\includegraphics[width=5.0in,angle=0]{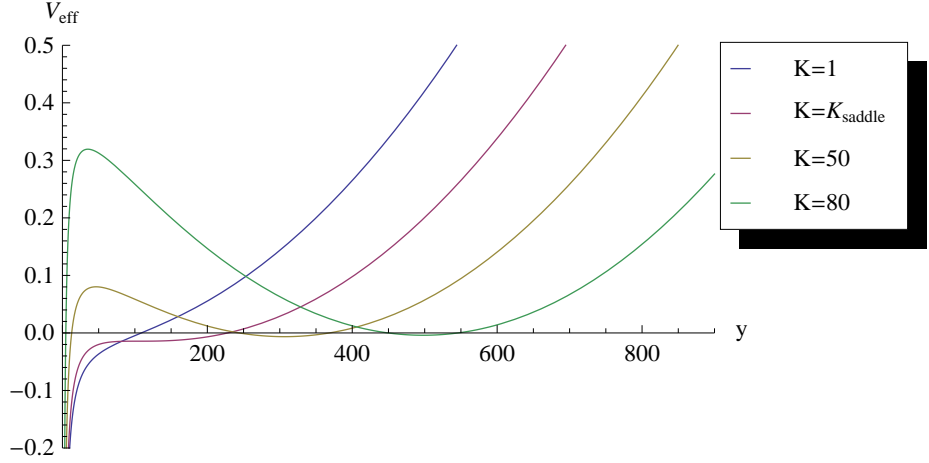}
\caption{$He^+$ ion: effective potential $V_{eff}$ (\ref{Veff}) at $K=1,K_{saddle},50,80$. $K_{saddle}\approx25.13$ and
$B = B_0 = 3.7598 \times 10^{10} \,G$\,. (c.f. \cite{ET-genq:2013})}
\label{VB1}
\end{center}
\end{figure}

The term $V_{eff}$ (\ref{VeffTot}) is gauge invariant, i.e. it does not contain the vector potential, and for any value of $K$ has a minimum at $x=y=0$ which corresponds to the Coulomb singularity. It can be called
the Coulomb minimum. For certain values of $K$ larger than some critical Pseudomomentum $K_{saddle}$ another minimum can occur.
It is located along the line perpendicular to the $x$-direction. In this direction, at $x=0$, $V_{eff}$ reads
\begin{equation}
  V_{eff}(0,y) \ =\  \frac{ {(K - B\,e\,\mu \,y)}^2}{2\,M}  - \frac{2\,e^2}{\mid y\mid } \ .
\label{Veff}
\end{equation}
and the position $y_0$ of minimum is given by a solution of the cubic equation
\begin{equation}
\begin{aligned}
y_0^3-\frac{K}{e\,\mu \,B}\,y_0^2-\text{sign}[y_0]\frac{2\,M}{\mu^2\,B^2}=0\ .
\label{Vy}
\end{aligned}
\end{equation}
All three solutions of (\ref{Vy}) are real if
\begin{equation}
\begin{aligned}
  K\ \geq \  K_{saddle}\ \equiv \ {\bigg(\frac{27}{4}e^3\,\mu \,B\,M\bigg)}^{\frac{1}{3}}\ .
\label{Psadd}
\end{aligned}
\end{equation}
At $K=K_{saddle}$ the eq.(\ref{Vy}) has a double zero which corresponds to the appearance of the saddle point in (\ref{Veff}).
It is located at $y_{saddle} = {(\frac{4\,M}{\mu^2\,B^2})}^{\frac{1}{3}}$. For $K > K_{saddle}$, the potential (\ref{Veff}) has two minima,
(see Fig.\ref{VB1}). For fixed $B$ in the limit $K \rar \infty$ we can easily obtain from (\ref{Vy}) the expression
\begin{equation}
\begin{aligned}
       y_{0,min} \approx \frac{K}{e\,\mu \,B} - \frac{2\,e^2\,M}{K^2} + \ldots \ ,
\label{min}
\end{aligned}
\end{equation}
therefore, the minimum $y_{0,min}$ grows linearly at large $K$ and $V_{eff}(0,y_{0,min})$ tends to zero from below as $-\frac{2\,B\,e^3\,\mu}{K}$.
Similarly, the position of the maximum
\begin{equation}
\begin{aligned}
  y_{0, max}\approx\sqrt{\frac{e\,M}{\mu \,B\,K}} + \frac{e^2\,M}{K^2}\ +\ \ldots
\label{max}\ ,
\end{aligned}
\end{equation}
thus, $y_{0,max}\rar 0$ with grows of $K$ and $V_{eff}(0,y_{0,max})\rar \infty$ as $K^2$.
The behavior of the barrier height $\De V_{eff}=V_{eff}(0,y_{0,max})-V_{eff}(0,y_{0,min})$ at large $K$
is given by the expansion
\begin{equation}
\begin{aligned}
 \De V \ = \ \frac{K^2}{2\,M} - \sqrt{\frac{4\,B\,e^3\,\mu \,K}{M}} + \frac{3\,B\,e^3\,\mu}{2\,K}    + \ldots
\label{DV}
\end{aligned}
\end{equation}

For $K=0$ the second minimum in $V_{eff}$ (\ref{Veff}) does not exist and the potential possesses azimuthal symmetry.
In this case, the symmetric gauge emerges naturally as the most convenient choice. The
convenience is related with the fact that for this gauge the ground state eigenfunction is real.
For $K\neq 0$ the azimuthal symmetry is broken, consequently, the most convenient choice of the gauge to treat the problem is no longer evident.
A question can be posed: in what gauge the ground state eigenfunction is real? In such a gauge the trial function for the ground state can be searched among real functions.
This strategy was realized in \cite{ET-genq:2013} and \cite{PRepts}.

Since we are going to use an approximate method for solving the Schr\"{o}dinger equation with the Hamiltonian (\ref{HKa}), a quality
of the approximation of ground state function can depend on the gauge. In particular, one can ask whether one can find a gauge for which a given trial
function leads to minimal variational energy. Such a gauge (if found) can be called optimal for a chosen trial function.

To this end, it is convenient to introduce a gauge transformation
\begin{equation}
\label{UU}
  {U}\ =\ e^{i\,B\,q_{\rm{w}}\,(1-\xi)(1-\nu)\,{y}_0 \, x}\ ,
\end{equation}
where
\begin{equation}
\label{cengaug}
{y}_{0}=-\frac{d\,K}{e\,\mu \,B}
\end{equation}
and $\nu,\,d$ are parameters. The gauge transformed Hamiltonian
(\ref{HKa}) takes the form
\begin{equation}
\begin{aligned}
  {h}_{d,\,\nu} \ \equiv\  U^{-1}\,{h}\, U \  =& \ \frac{ (- \partial^2_{{}_Y} - 2\,i\,B\,e\,\mu \,x\,\partial_{{}_Y} + e^2\,B^2\,Y^2 +2\,e\,B\,Y\,(K - B\,e\,\mu \,y))}{2\,M}
\\ & \ +      \frac{{({\mathbf {\hat p}}-q_\text{w}\,{\mathbf A_{(\boldsymbol \rho-{ \boldsymbol \rho}_0)}})}^2}{2\,m_{r}} +V_{eff}                      \ ,
\label{HP}
\end{aligned}
\end{equation}

where ${ \boldsymbol \rho}_0=y_0(1-\nu)\,\hat{\mathbf y}$. This transformation implies that we consider now the Schr\"{o}dinger
equation in a linear gauge for which the position of the gauge center,
where $\mathbf A(x,y)=0$, is located at

\begin{equation}
x=0 \,, \quad y = \frac{d\,K}{e\,\mu \,B}(1-\nu) \ .
\label{ygauge}
\end{equation}

For $K > K_{saddle}$ we expect the gauge center to be localized on the line $x=0$, between the origin $y=0$ and the second
minimum $y=y_{0,min}$ of $V_{eff}$, (see (\ref{min})).
Thus, the vector potential can be considered as a variational function and can be chosen by a procedure
of minimization as it was proposed in \cite{PRepts} and realized in \cite{ET-genq:2013} (see also for discussion \cite{Vincke}).
For a moving neutral system, the case $d=0$ has been used in the past to study the so-called centered states with wavefunction peaked at the Coulomb minimum \cite{Burkova}.
While for the so-called "decentered" states it seems natural to consider $d=1$. The eigenvalue problem
\begin{equation}
{h}_{d,\nu}\,\chi_{{}_{K}}= E \,\chi_{{}_{K}} \ ,
\label{hxi}
\end{equation}
where $\chi_{{}_{ K}}(\boldsymbol \rho,\,Y)\ =\ e^{-i\,B\,q_{\rm{w}}\,(1-\xi)(1-\nu)\,{y}_0 \, x} \,\psi_{{}_{K}}(\boldsymbol \rho,\,Y) $, is the central object of our
study hereafter. For convenience, in the calculations we used the so called shifted representation, $\chi_{{}_{K}}(\boldsymbol \rho,\,Y)\rightarrow
\chi_{{}_{K}}(\boldsymbol \rho+y_0\,\hat{\mathbf y},\,Y)$.

\bigskip

\subsection{Asymptotics.}

\vspace{0.2cm}

If we put $\chi_{{}_{K}} = \text{e}^{-\varphi}$ and $\xi=\frac{1}{2}$ in (\ref{hxi}),
one can construct the WKB-expansion at large $\rho = \sqrt{x^2+y^2}$ for the phase $\varphi$.
The leading term at $\rho \rar \infty$ is given by
\begin{equation}
     \varphi \ =\  \frac{B}{2}{\bigg(\frac{m_r}{M}e^2\,\mu^2 +\frac{q_{\rm{w}}^2}{4} \bigg)}^{\frac{1}{2}}\, \rho^2  +  O(\rho) \ .
\label{large}
\end{equation}

Similarly, at $Y \rar \infty$

\begin{equation}
\varphi \ =\  \frac{e\,B}{2}\,Y^2  +  O(Y) \ .
\label{largeY}
\end{equation}

In the limit $\rho \rar 0$ we obtain

\begin{equation}
\varphi \ =\  -4\,m_r\,e^2\,\rho +  O(\rho^2) \ .
\label{small}
\end{equation}

Assuming the condition (\ref{Psadd}) is fulfilled, the potential (\ref{Veff})
has the second minimum at $y_{0,min}\neq 0$. Hence, the double Taylor expansion of the phase at $x = 0, y= y_{0,min}$ has the form
\begin{equation}
\begin{aligned}
\varphi &\ =\  \al_0+\al_2\,{(y-y_{0,min})}^2  + \al_3\,{(y-y_{0,min})}^3 + \al_4\,{(y-y_{0,min})}^4 +  \ldots
\\ &\ + \beta_2\,x^2+\beta_4\,x^4 + \ldots + \gamma_3\,(y-y_{0,min})\,x^2 +   \ldots
\label{y0}
\end{aligned}
\end{equation}
where $\al$\textquoteright s, $\beta$\textquoteright s and $\gamma$\textquoteright s are constants.

\bigskip

\subsection{Approximations}

Following the prescription formulated in \cite{Turbiner:1988-2010} we make
an interpolation between WKB-expansion (\ref{large}),(\ref{largeY}) and the perturbative expansion (\ref{small}),(\ref{y0}) and construct a trial function for
the ground state of (\ref{hxi}) in a form of a product
\begin{equation}
\chi_{{}_{K}}(\boldsymbol \rho,\,Y)  \ = \ e^{-a^2\,Y^2+b\,Y\,y-\beta^2\,y^2}\ \zeta(\boldsymbol \rho)\,\ ,
\label{interp}
\end{equation}

where

\begin{equation}
\begin{aligned}
& \zeta(\boldsymbol \rho) \ =\  C_1 \,e^{-\phi_c} + C_2\,e^{-\phi_m}
\\ &    \phi_c = \ \frac{A_0 + A_1\,{\rho} + A_2\,{y}\,{\rho} +  A_3^2\,{\rho^3}
}{   \sqrt{ 1 + A_4\,{y} + A_5^2 \,{\rho^2}}   } - \frac{\alpha}{2}\,\log(1+A_4\,{y} + A_5^2\,{\rho^2})
\\  &  \phi_m = \ \frac{D_0 + D_1\,{{x}^2}+D_2\,{{\tilde y}^2} +D_3^2\,{\varrho^4}}{
  \sqrt{1 + D_4\,{{x}^2} +  D_5\,{{\tilde y}^2}  +  D_{6}^2\,{\varrho^4}}}
\label{phis}
\end{aligned}
\end{equation}

with $\tilde y = y-y_0$, $\varrho^2 = {x}^2+{\tilde y}^2$ and $a,\,b,\,\beta,\,A^\prime s,\,C^\prime s,D^\prime s,\,\alpha$ variational parameters.
Supposedly, they should behave smoothly as a function of a magnetic field.

As mentioned above this problem has some similarity with that of the moving neutral system for which a physically adequate trial function is $\zeta(\boldsymbol \rho)$.
The difference comes from the first term in r.h.s. of (\ref{HKa}), the contribution of this term is encoded in the factor $e^{-a^2\,Y^2+b\,Y\,y-\beta^2\,y^2}$ in (\ref{interp}).
The parameter $b$ measures the coupling between CM and relative variables, $b=0$ corresponds to the adiabatic approximation.

\section{Results}

\hskip 1cm
We carried out a variational study of the two-body charged system on a plane moving across a magnetic field.
The main emphasis is to explore stability of the system, thus, studying the ground state. For the case of
$He^+$ ion, the energy for several magnetic fields $0 < B < 100$\,a.u. and values of Pseudomomentum $0 \leq K < 1000$ a.u. is presented in Table \ref{Table1}.
The energy grows monotonically and rather sharp as a function of a magnetic field for fixed Pseudomomentum but
at much slow pace as a function of Pseudomomentum for a fixed magnetic field.
It is worth noting that for fixed $B$ the energy $E$ as a function of $K$ tends asymptotically to the ground state energy of two non-interacting charges in a magnetic field.

\begin{figure}[htp]
\centering
\subfigure[]{\includegraphics[width=3.5in,angle=-90]{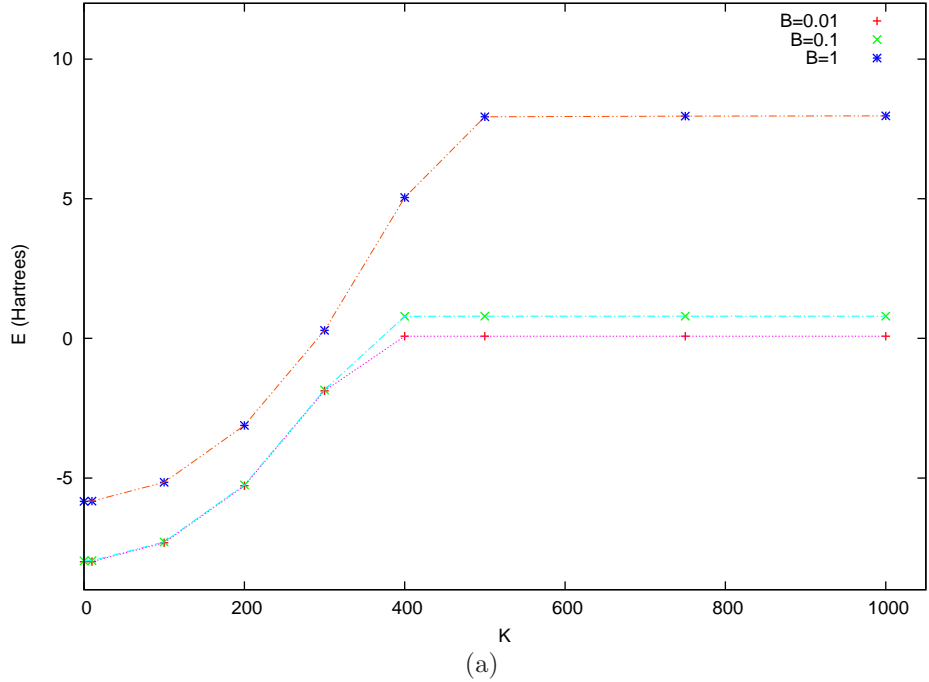}} \qquad \subfigure[]{\includegraphics[width=3.5in,angle=-90]{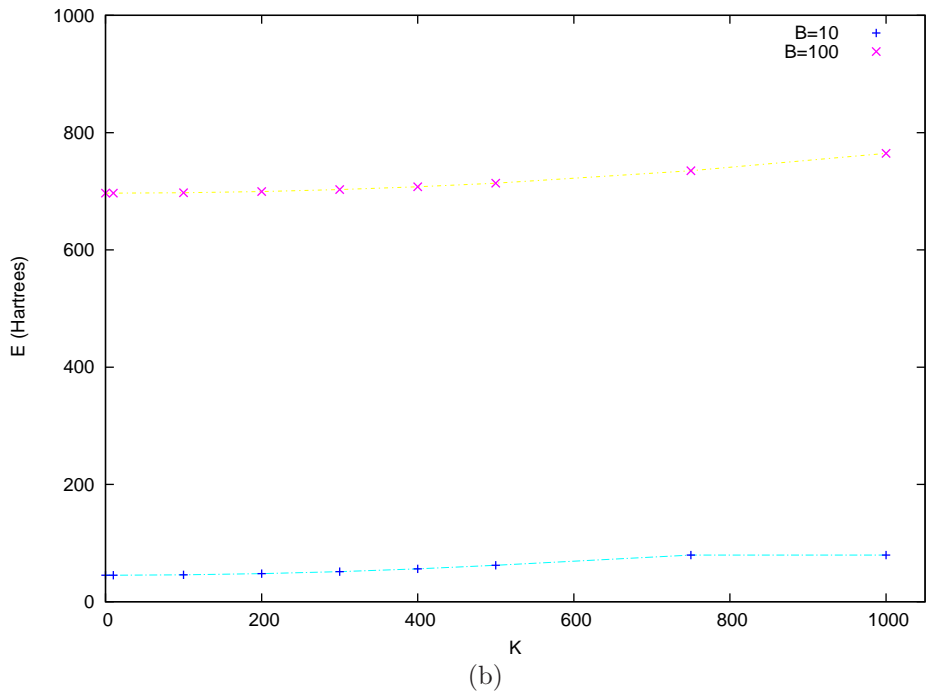} }
\caption{The energy (in Hartrees) of ground state {\it vs} Pseudomomentum $K$. (a) $E$ \textit{vs} $K$
for magnetic fields $B=0.01,\,0.1,\,1$. (b) $E$ \textit{vs} $K$ for magnetic fields $B=10,\,100$\,a.u.; magnetic field $B$ and Pseudomomentum $K$
in effective atomic units, $B_0 = 3.7598 \times 10^{10} \,G$. (c.f. \cite{ET-genq:2013} Fig. 2)
 }
\label{E0}
\end{figure}

\hskip 1cm
After making a minimization, one can see the appearance of a sharp change in the behavior of parameters in (\ref{phis}) as a function of $K$.
It is related with a fact of the existence of a certain
critical Pseudomomentum $K_c > K_{saddle}$ such that for $K < K_c$ the optimal linear parameters
$C_1 \approx 1,\,C_2 \approx 0$ the wavefunction has a peak near the Coulomb singularity (centered state). At $K > K_c$
the situation gets opposite: the parameters $C_1 \approx 0,\,C_2 \approx 1$ and the wavefunction is peaked near the second well of (\ref{Veff}),
see Fig.\ref{VB1} (we call this well the {\it magnetic} well) which corresponds to a decentered state. A similar phenomenon appears in the case of a neutral
system in 3D \cite{Burkova} and 2D \cite{Lozovik:2002}. The existence of such a change in the behavior of parameters results also in a specific behavior of energy dependence
(see for example Fig.\ref{E0}) and mean interparticle separation {\it vs.} the Pseudomomentum.
From physical point of view at $K = K_c$ the effective depth of the Coulomb well and that of the magnetic well get equal. If $K < K_c$
the effective depth of the Coulomb well is larger (or much larger depending on a magnetic field strength) than one of the magnetic well, the system prefers
to stay at the Coulomb well. If $K > K_c$ the effective depth of the Coulomb well is smaller (or much smaller depending on a magnetic field strength) than one
of the magnetic well, the system prefers to stay at the magnetic well. For all studied magnetic fields the barrier between wells is very large,
the probability of tunneling from one well to the other is very small. Hence, the energy behavior {\it vs.} CM  Pseudomomentum $K$ is defined by one well or another,
it is close to classical behavior. Thus, the presence of the second minimum in the effective potential can be neglected.

\begin{figure}[htp]
\begin{center}
\includegraphics[width=4.0in,angle=-90]{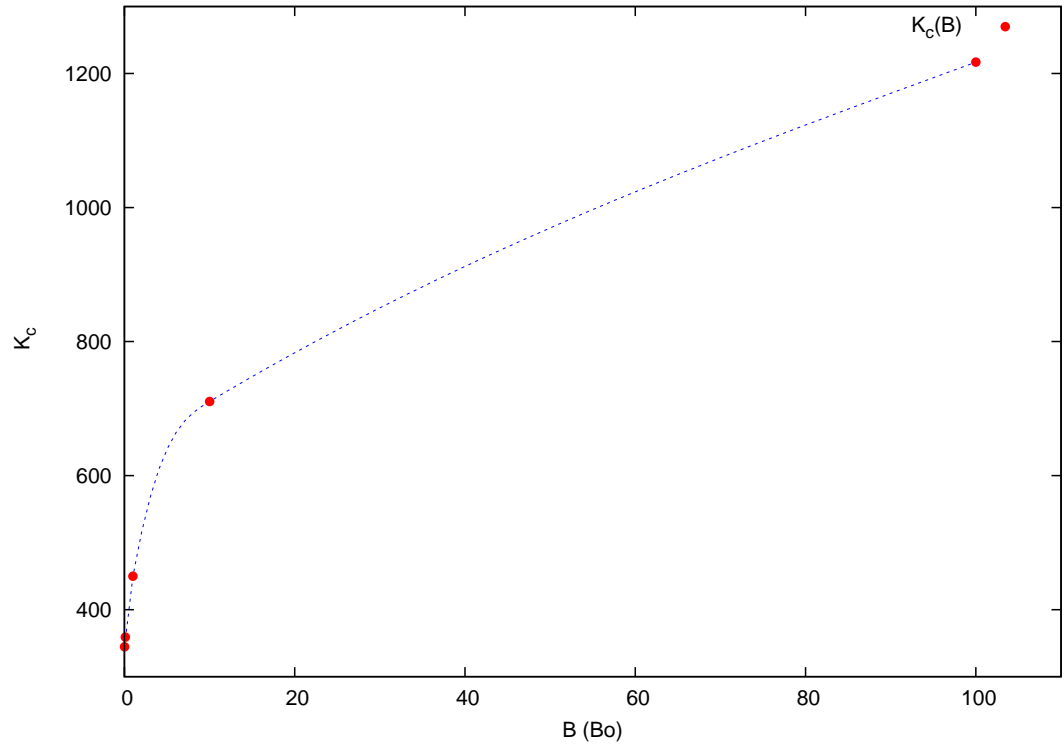}
\caption{Critical Pseudomomentum $K_c$ \textit{vs} $B$\,, (c.f. \cite{ET-genq:2013} Fig. 3) }
\label{Pcr:fig.3}
\end{center}
\end{figure}

\hskip 1cm
The results of calculations show that the optimal gauge parameter $\xi$ for all values of magnetic field considered always
corresponds to symmetric gauge $\xi=\frac{1}{2}$. The behavior of $K_c$ as a function of magnetic field is presented in Fig. \ref{Pcr:fig.3} where $K_c$ grows with an increase of $B$.
The evolution of the gauge center parameters ($d,\,\nu$), see ((\ref{UU})-(\ref{ygauge})), {\it vs.} CM Pseudomomentum $K$ is shown in Tables \ref{Table2}-\ref{Table2a},
respectively. For $K < K_c$ the parameter $d$ is very small, it varies within $[0 - 5\times10^{-5}]$ for magnetic field range $0.01 - 100$\,a.u. while for
$K > K_c$ it is close to 1, it varies within $[1 - 0.996]$ for magnetic field range $0.01-10$\,a.u. At $B<1$ and for any $K$ considered the parameter $\nu \sim 0$.
At fixed $B\eqslantgtr1$ the parameter $\nu=\nu(B)\neq 0$ for $K<K_c$ and almost constant for $K   \gtrsim K_c$.
Thus, the gauge as a function of $K$ changes from the symmetric gauge centered at the Coulomb well (the singularity of (\ref{Veff})) to the symmetric gauge 
but centered at the magnetic well (the minimun of (\ref{Veff}) for $K>K_c$).
In turn, the parameter $d$, which mainly determines the value of the gauge center, remains almost equal to $0$ up to $K = K_c$ (which means the gauge center
coincides with a position of the Coulomb singularity, then sharply jumps to a value close to $1$ (gauge center coincides with a position of the minimum of magnetic well),
displaying a behavior which looks like a phase transition. But it is not a phase transition: the energy changes sharply but smoothly. For $K = K_c$ the gauge parameters are
$d \sim 0.5,\,\nu \sim 0$. There exists a certain domain of transition from one regime to another. Overall
situation looks very similar to that for the $H_2^+$ molecular ion in a magnetic field in inclined configuration \cite{PRepts}.

\hskip 1cm
In order to illustrate the transition from a centered state to a decentered one we have calculated,
using the trial function (\ref{interp}) with optimal parameters, the expectation value of the relative coordinate
$\langle \rho \rangle$, see Table \ref{exprho}. At weak magnetic fields $B$,
the transition is very sharp, becoming even more pronounced with a magnetic field decrease.
For all studied magnetic fields and Pseudomomentum both $\langle \rho \rangle$ and $\langle Y \rangle$ are finite.
Furthermore, the trial function (\ref{interp}) remains normalizable. It indicates the stability and boundedness of the $He^+$ ion in magnetic field.

\hskip 1cm
To complete the study we show in Tables VIII - XIV (see Supplementary Materials) the
nonlinear parameters $\beta$, $A$'s, $D$'s and $\al$ of the trial function (\ref{phis}) as
a function of the magnetic field strength for the optimal configuration. For all considered values of Pseudomomentum the parameter
$A_1\approx 4$. A deviation $|A_1-4|$ measures (anti)-screening of the electric charge due to the presence of a magnetic field.
The optimal value of energy corresponds to $y_0 \approx y_{0,min}$. Similarly, for all considered values of Pseudomomentum the parameter
$a\approx0.283,\,0.8943,\,2.828,\,8.942,\,28.277$ at $B=0.01,\,0.1,\,1,\,10,\,100$, respectively. A deviation $|1 - \frac{1}{a}\sqrt{\frac{e\,B}{2}}|\lesssim10^{-4}$ measures the
correctness of the asymptotic behavior of the trial function, see (\ref{largeY}).

\hskip 1cm
In Table \ref{Tableb} the evolution of parameter $b$ is presented, it shows that
the coupling between CM and relative variables is a non-decreasing function of magnetic field. For fixed $B$ the behavior is different, at $K<K_c$ the optimal value
of $b$ is a decreasing function (changing from a positive value to a negative one) of Pseudomomentum while for $K>K_c$ the parameter $b>0$ is almost constant. Because there is no
separation of variables in the problem, $b$ is never zero.
Clearly, the parameter $b$ will play an important role either for $B\rightarrow \infty$ or for systems with $m_r \sim 0.5$.

\hskip 1cm
Our variational results are checked on agreement with results obtained with other methods.
We used the Lagrange mesh method (see \cite{Baye} and references therein) to obtain
the ground energy for $B=0.1$\,a.u. and different Pseudomomentum $K$, see Table \ref{TabMesh}.
For all studied values of $K$ the variational energy is in agreement with Lagrange mesh calculations in not less than 5 s.d. .
It is interesting to check the accuracy of the Born-Oppenheimer approximation (taking $m_2\rightarrow \infty$) as well. The results are
presented in Table \ref{Eper}. The relative difference in energy due to the finite mass effects are of order $\approx 10^{-4}$ for all $B$, as expected.

\clearpage

\begin{center}
\textbf{$He^+$ ion. Energy $E$}
\end{center}
\setlength{\tabcolsep}{11.0pt}
\setlength{\extrarowheight}{1.0pt}
\begin{table}[th]
\begin{center}
\begin{tabular}{|c||c|c|c|c|c|}
\hline
\\[-23pt]
$K$   & \multicolumn{5}{c|}{ $E$ } \\
\hline               & \multicolumn{5}{l|}{\hspace{0.0cm} $B=0.01$ \hspace{0.6cm} $B = 0.1$ \hspace{0.6cm}
$B=1$ \hspace{0.6cm} $B =10$ \hspace{0.6cm} $B=100$ }\\
\hline
\hline
$0$                  &   $-7.9986$    &   $-7.9690$         &   $-5.8365$   &   $45.226$     &   $696.89$
\\[3pt]
\hline
$10$                  &   $-7.9918$    &   $-7.9622$         &   $-5.8297$   &   $45.232$     &   $696.90$
\\[3pt]
\hline
$100$                 &   $-7.3184$    &   $-7.2888$         &   $-5.1563$   &   $45.905$     &   $697.57$
\\[3pt]
\hline
$200$                 &   $-5.2778$    &   $-5.2482$         &   $-3.1159$   &   $47.944$     &   $699.60$
\\[3pt]
\hline
$300$                 &   $-1.8767$    &   $-1.8472 $          &   $0.2846$   &   $51.342$     &   $702.99$
\\[3pt]
\hline
$400$                &   $0.07920^d$  &   $0.7920^d$           &   $5.0454$   &   $56.098$     &   $707.74$
\\[3pt]
\hline
$500$                &   $0.07936^d$  &   $0.7936^d$         &   $7.9359^d$   &   $62.214$     &   $713.84$
\\[3pt]
\hline
$750$                &   $0.07957^d$  &   $0.7957^d$         &   $7.9573^d$   &   $79.572^d$     &   $735.02$
\\[3pt]
\hline
$1000$                &   $0.07968^d$  &   $0.7968^d$         &   $7.9680^d$   &   $79.680^d$     &   $764.66$
\\[3pt]
\hline
\end{tabular}
\caption{\small Ground state energy $E$ in Hartrees (see (\ref{hxi})); magnetic field $B$ and Pseudomomentum $K$ in effective atomic units, $B_0 = 3.7598 \times 10^{10} $\,G.
, $K / \hbar = 5.5 \times 10^{12}\, { \rm cm}^{-1}$, respectively. Energies corresponding to decentered states marked by ${}^d$\,(see text).}
\label{Table1}
\end{center}
\end{table}

\clearpage

\begin{center}
\textbf{$He^+$ ion. Parameter $d$}
\end{center}
\setlength{\tabcolsep}{22.0pt}
\setlength{\extrarowheight}{1.0pt}
\begin{table}[th]
\begin{center}
\begin{tabular}{|c||c|c|c|c|c|}
\hline
\\[-22pt]
$K$   & \multicolumn{5}{c|}{ $d$ } \\
\hline                 & \multicolumn{5}{l|}{\hspace{0cm} $B=0.01$ \hspace{0.5cm} $B = 0.1$ \hspace{1.2cm}
$B=1$ \hspace{1.5cm} $B =10$ \hspace{2.0cm} $B=100$ }\\
\hline
\hline
$K<K_c$                     &   $0.0$       &   $0.0$            &   $-2.0\times 10^{-6}$   &   $-5\times 10^{-6}$     &   $5.0\times 10^{-5}$
\\[3pt]
\hline
$K>K_c$                     &   $1.0$       &   $0.9998$         &   $0.999$                &   $0.996$                  &   $-$
\\[3pt]
\hline
\end{tabular}
\caption{\small Optimal gauge parameter $d$ (see (\ref{ygauge})) for different magnetic fields and pseudomomenta; magnetic field
in effective atomic units, $B_0 = 3.7598 \times 10^{10}$\,G.}
\label{Table2}
\end{center}
\end{table}

\begin{center}
\textbf{$He^+$ ion. Parameter $\nu$}
\end{center}
\setlength{\tabcolsep}{24.0pt}
\setlength{\extrarowheight}{1.0pt}
\begin{table}[th]
\begin{center}
\begin{tabular}{|c||c|c|c|}
\hline
\\[-23pt]
$K$   & \multicolumn{3}{c|}{ $\nu$ } \\
\hline                 & \multicolumn{3}{l|}{  $B =1$  \hspace{1.0cm}  $B=10$ \hspace{1.2cm} $B =100$ }\\
\hline
\hline
$0$                    &    $0$             &   $0$           &   $0$
\\[3pt]
\hline
$10$                   &    $-27.292 $     &   $-22.121$      &   $79.715$
\\[3pt]
\hline
$100$                  &    $-131.66 $     &   $-68.226$      &   $135.59$
\\[3pt]
\hline
$200$                  &    $-305.13$      &   $-316.21$      &   $135.60$
\\[3pt]
\hline
$300$                  &    $-638.94 $     &   $-570.81$      &   $135.79$
\\[3pt]
\hline
$400$                  &    $-638.94 $     &   $-746.60$      &   $95.506$
\\[3pt]
\hline
$500$                  &    $0$            &   $-750.29$      &   $34.637$
\\[3pt]
\hline
$750$                  &    $0$            &   $0$            &   $34.527$
\\[3pt]
\hline
$1000$                 &    $0$            &   $0$            &   $5.1179$
\\[3pt]
\hline
\end{tabular}
\caption{\small Ground state. Parameter $\nu$ in (\ref{ygauge}). At $B=0.01,\,0.1$, parameter $\nu \sim 0$ for all $K$ considered.
$B_0 = 3.7598 \times 10^{10} \,G$.}
\label{Table2a}
\end{center}
\end{table}

\clearpage

\clearpage

\begin{center}
\textbf{$He^+$ ion.  Expectation value $\langle \rho \rangle$}
\end{center}
\setlength{\tabcolsep}{26.0pt}
\setlength{\extrarowheight}{1.0pt}
\begin{table}[th]
\begin{center}
\begin{tabular}{|c||c|c|c|}
\hline
\\[-23pt]
$K$   & \multicolumn{3}{c|}{ $\langle \rho \rangle$ } \\
\hline                 & \multicolumn{3}{l|}{\hspace{0.0cm} $B=0.01$ \hspace{0.8cm} $B = 0.1$ \hspace{0.8cm}
$B=1$  }\\
\hline
\hline
$0$                     &   $0.250$       &   $0.248$        &   $0.198$
\\[3pt]
\hline
$100$                   &   $0.250$      &   $0.248$         &   $0.198$
\\[3pt]
\hline
$300$                   &   $0.250$     &   $0.248$         &   $0.198$
\\[3pt]
\hline
$400$                   &   $2500^d$     &   $249^d$         &   $0.198$
\\[3pt]
\hline
$500$                   &   $3125^d$     &   $312^d$         &   $31^d$
\\[3pt]
\hline
$750$                   &   $4688^d$     &   $468^d$         &   $46^d$
\\[3pt]
\hline
$1000$                   &   $6250^d$     &   $625^d$         &   $62^d$
\\[3pt]
\hline
\end{tabular}
\caption{\small Expectation value $\langle \rho \rangle$; magnetic field in effective atomic units, $B_0 = 3.7598 \times 10^{10}\,G$.
At $B=10, 100$\,a.u.,\ $\langle \rho \rangle \approx 0.087, 0.018$, respectively. Data corresponding to decentered states marked by ${}^d$ \,(see text).
For $K<K_c$, $\langle \rho \rangle \approx 0 $, while for $K>K_c$,
$\langle \rho \rangle \sim \langle y\rangle \sim y_{0,min}\propto K/B$\ .}
\label{exprho}
\end{center}
\end{table}

\clearpage

\begin{center}
\textbf{$He^+$ ion. Parameter $b$}
\end{center}
\setlength{\tabcolsep}{24.0pt}
\setlength{\extrarowheight}{1.0pt}
\begin{table}[th]
\begin{center}
\begin{tabular}{|c||c|c|c|}
\hline
 $K$      & \multicolumn{3}{c|}{    $B=1$   \hspace{1.5cm}     $B=10$        \hspace{1.5cm}     $B=100$   } \\[2pt]
\hline  & \multicolumn{3}{c|}{      $a=2.828$   \hspace{0.9cm}     $a=8.942$        \hspace{0.9cm}     $a=28.277$         }\\[2pt]
\hline
\hline
\hline
$0$                    &    $0.002$   &   $0.0349$           &   $0.4079$
\\[3pt]
\hline
$10$                   &    $0.0019$    &   $0.0349$        &   $0.4021$
\\[3pt]
\hline
$100$                  &    $0.0014$     &   $0.0312$      &   $0.3983$
\\[3pt]
\hline
$200$                  &    $0.0003$      &   $0.0299$      &   $0.3876$
\\[3pt]
\hline
$300$                  &    $-0.0013$     &   $0.0094$      &   $0.3475$
\\[3pt]
\hline
$400$                  &    $-0.0040$     &   $-0.0033$      &   $0.2973$
\\[3pt]
\hline
$500$                  &    $0.0043^d$      &   $-0.0217$      &   $0.2395$
\\[3pt]
\hline
$750$                  &    $0.0043^d$      &   $0.0435^d$            &   $0.0371$
\\[3pt]
\hline
$1000$                 &    $ 0.0043^d$     &   $0.0435^d$            &   $-0.4303$
\\[2pt]
\hline
\end{tabular}
\caption{\small Ground state. Parameters $a,\,b$ in (\ref{interp}). At $B=0.01,\,0.1$, parameter $b\sim10^{-4}$ and $a=0.283,\,0.8943$ respectively, for all $K$ considered.
Data corresponding to decentered states marked by ${}^d$ \,(see text).
$B_0 = 3.7598 \times 10^{10} \,G$.}
\label{Tableb}
\end{center}
\end{table}

\clearpage

\setlength{\tabcolsep}{25.0pt}
\setlength{\extrarowheight}{0.0pt}
\begin{table}[th]
\begin{center}
\begin{tabular}{|c||c|c|}
\hline
\\[-21pt]
$B$   & \multicolumn{2}{c|}{ $E$ } \\
\hline                 & \multicolumn{2}{c|}{$E_0$ \hspace{1.9cm} $E_0^\infty$ }\\
\hline
\hline
$0.01$                &   $-7.9986$       &   $-7.9997$
\\[3pt]
\hline
$0.1$                 &   $-7.9690$        &   $-7.9702$
\\[3pt]
\hline
$1$                   &   $-5.8365$        &   $-5.8393$
\\[3pt]
\hline
$10$                  &   $45.226$         &   $45.203$
\\[3pt]
\hline
$100$                 &   $696.89$         &   $696.67$
\\[3pt]
\hline
\end{tabular}
\caption{\small Ground state energy in Hartrees. $E_0$ from Table \ref{Table1} and $E_0^\infty$ ($m_2\rightarrow\infty$).
Magnetic field in effective atomic units, $B_0 = 3.7598 \times 10^{10} \,G$ .}
\label{Eper}
\end{center}
\end{table}

\setlength{\tabcolsep}{25.0pt}
\setlength{\extrarowheight}{0.0pt}
\begin{table}[th]
\begin{center}
\begin{tabular}{|c||c|c|}
\hline
\\[-21pt]
$K$   & \multicolumn{2}{c|}{ $E$ } \\
\hline                 & \multicolumn{2}{c|}{$E_0$ \hspace{2.6cm} $E^{mesh}_{0}$ }\\
\hline
\hline
$0$                      &   $-7.969046$            &   $-7.969047$
\\[3pt]
\hline
$10$                     &   $-7.962244$            &   $-7.962245$
\\[3pt]
\hline
$100$                    &   $-7.288839$            &   $-7.288839$
\\[3pt]
\hline
$300$                    &   $-1.847176$            &   $-1.847177$
\\[3pt]
\hline
$500$                    &   $0.793600$             &   $0.793600$
\\[3pt]
\hline
$1000$                   &   $0.796800$             &   $ 0.796800$
\\[3pt]
\hline
\end{tabular}
\caption{\small Ground state energy at $B=\frac{1}{10}$. $E_0$ from the present study and $E_0^{mesh}$ obtained with the Lagrange mesh method
(parameters $d,\nu$ taken from the corresponding variational results based on (\ref{interp})).
Magnetic field in effective atomic units, $B_0 = 3.7598 \times 10^{10} \,G$.}
\label{TabMesh}
\end{center}
\end{table}

\clearpage

\section{Conclusions}

\hskip 1cm
Summarizing, for the two-dimensional
$He^+$ ion in a constant magnetic field partial factorization of eigenfunctions (see (\ref{psik})) allows us to reduce the problem to one with three degrees of freedom. For this reduced problem
we want to state that a simple uniform approximation of the ground
state eigenfunction is constructed. It
manifests an approximate solution of the problem. The key element of the procedure is to make an interpolation between the WKB expansion
at large distances and perturbation series at small distances both for the phase of the wavefunction; in other words,
to find an approximate solution for the corresponding eikonal equation. In general, the separation of
variables helps us to solve this problem easily.
In our case of non-separability of variables the WKB expansion of a solution of the eikonal equation cannot
be constructed in a unified way, since it depends on the way how we approach  to infinity. However, a reasonable approximation
of the first dominant growing terms of the WKB expansion of the phase seems sufficient to construct the interpolation between
large and small distances giving rather high accuracy results.

\hskip 1cm
It was demonstrated that for all magnetic fields and all values of Pseudomomentum the system is bounded.
Its energy grows with magnetic field strength increase as well as Pseudomomentum increase. For fixed magnetic field $B$, the energy behavior demonstrates
a sharp change for a certain value of CM Pseudomomentum $K_c (B)$. It seems it can be used to measure the magnetic field strength.
This effect was already mentioned in three-dimensional Hydrogen atom moving across
magnetic field \cite{Potekhin}.

\hskip 1cm
In the Born-Oppenheimer approximation ($m_2\rightarrow \infty$), a curious fact that the Hamiltonian (\ref{H}) possesses the hidden
algebra $sl_2(\Re)$ is worth mentioning. It can be immediately seen-making a gauge rotation of the Hamiltonian (\ref{H}) in symmetric gauge ($\xi=\frac{1}{2}$)
with the gauge factor $e^{-\frac{e\,B}{4}\rho_1^2}$ . We obtain the operator which is
in the universal enveloping algebra of $sl_2(\Re)$ (see e.g. \cite{Turbiner:1994}).
Hence, for specific values of a magnetic field $B$ the algebra $sl_2(\Re)$ appears in finite-dimensional
representation and the problem admits analytical solutions (details will be given elsewhere).

\begin{acknowledgments}
  The author would like to thank J. C. L\'opez Vieyra and H. Olivares Pilon for their interest in the present work, helpful discussions and important
  assistance with computer calculations. I am especially grateful to A. V. Turbiner, who initiated this work and gave priceless advice during its realization.
  This work was supported in part by the University Program FENOMEC, and by the PAPIIT
  grant {\bf IN109512} and CONACyT grant {\bf 166189}~(Mexico). The author is supported by CONACyT project for postdoctoral research.
\end{acknowledgments}

\end{document}